\documentclass{article}

\usepackage{arxiv}

\usepackage[utf8]{inputenc} % allow utf-8 input
\usepackage[T1]{fontenc}    % use 8-bit T1 fonts
\usepackage{hyperref}       % hyperlinks
\usepackage{url}            % simple URL typesetting
\usepackage{booktabs}       % professional-quality tables
\usepackage{amsfonts}       % blackboard math symbols
\usepackage{nicefrac}       % compact symbols for 1/2, etc.
\usepackage{microtype}      % microtypography
\usepackage{lipsum}

\usepackage{graphicx}% Include figure files
\usepackage{dcolumn}% Align table columns on decimal point
\usepackage{bm}% bold math

\usepackage{amsmath}

\usepackage{mathrsfs}
\usepackage{tikz}
\usepackage{amssymb}

\usepackage{graphicx}

\usepackage[numbers]{natbib}

\usepackage[algoruled,vlined]{algorithm2e}

\title{Framework for Converting Mechanistic Network Models to Probabilistic Models}

\author{
  Ravi Goyal\\
  Mathematica \\
  600 Alexander Park, Suite 100\\
  Princeton, NJ 08540\\
  \texttt{rgoyal@mathematica-mpr.com} \\
   \And
  JP Onnela\\
  Department of Biostatistics\\
  Harvard T.H. Chan School of Public Health\\
  655 Huntington Avenue \\
  Boston, MA 02115\\
  \texttt{onnela@hsph.harvard.edu} \\
}

\begin{document}
\maketitle

\begin{abstract}
There are two prominent paradigms to the modeling of networks: in the first, referred to as the mechanistic approach, one specifies a set of domain-specific mechanistic rules that are used to grow or evolve the network over time; in the second, referred to as the probabilistic approach, one describes a model that specifies the likelihood of observing a given network. Mechanistic models are scalable and, in select cases,  allow for analytical solutions for some of their properties, whereas probabilistic models have inferential tools available. Mechanistic models are appealing because they capture scientific processes that are hypothesized to be responsible for network generation. We introduce a generic framework for converting a mechanistic network model to a probabilistic network model. The proposed framework makes it possible to identify the essential network properties and their joint probability distribution for mechanistic network models, which enables addressing questions such as whether two mechanistic models generate networks with identical distributions of properties of interest, or whether a network property, such as clustering, is over- or under- represented in the generated networks compared to a reference model. The proposed framework is intended to bridge some of the gap that currently exists between mechanistic and probabilistic network models.
\end{abstract}

\keywords{networks \and mechanistic models \and probabilistic models}

\section{Introduction}

The utility of representing the structure of different complex systems as networks has been realized across disciplines from physics to public health \citep{MN10}. As the settings that give rise to these complex systems vary greatly, disciplines have developed specialized techniques for modeling and analyzing these systems. While there exists the potential of substantial synergy across the methodologies developed in different fields, limited tools currently exist to bridge these methodologies.  In this paper, we focus on bridging two of the primary techniques to generate simulated networks, mechanistic network models and probabilistic network models, with the goal of highlighting common ground between them. 

We use the term \emph{mechanistic network model} to refer to models that generate a network by repeatedly applying a collection of stochastic microscopic rules. Microscopic rules, denoted as $\gamma$, are typically domain specific, and they attempt to codify the essential organizing principles of the studied systems from the point of view of individual nodes. These rules can be deceptively simple and yet give rise to rich and complex network structure at the mesoscopic and macroscopic levels. Though mechanistic models do not explicitly specify a probability mass function (PMF) on a set of graphs, they do so implicitly; let $P_{\mathscr{G}}(G = g \vert \gamma)$ denote this implicit PMF, where $G$ is a random variable with support on $\mathscr{G}$ and $g \in \mathscr{G}$. 

We use the term \emph{probabilistic network model} to describe a model that specifies the likelihood of observing a given network, i.e., these models provide an explicit probability distribution on $\mathscr{G}$. Let $P_{\mathscr{G}}(G = g \vert \omega)$ denote this PMF, where $\omega$ represents functions or parameters necessary to specify the PMF; this formulation allows for the PMF to be parametric, semi-parametric, or non-parametric. The goal of specifying a probabilistic model is typically to estimate $\omega$ and the uncertainty associated with it from observed data. Then, given $\omega$, networks can be generated from the specified PMF, $P_{\mathscr{G}}(G = g \vert \omega)$, using a suitable computational procedure. 

Mechanistic and probabilistic network models each provide distinct insights and advantages to furthering our understanding of complex systems. Although these two methodologies appear to be quite distinct, they both have been applied in many of the same areas.  For example, in HIV prevention, networks generated based on mechanistic and probabilistic models have played a vital role in simulating HIV epidemics and, more recently, also simulating interventions to slow down the diffusion of the virus. Simulations results have been used to design large HIV prevention randomized trials and to prioritize interventions \citep{wang2014sample, granich2009universal, MKHMW09}. However, at present, there is very little dialogue between the two approaches. In this paper, we propose a framework for connecting network generation from the two model types. 

Mechanistic network models provide insight into how the network is formed at the level of individual decisions. However, as mechanistic rules typically lead to complex network structures, it is difficult to identify the PMF on $\mathscr{G}$ that a set of microscopic rules $\gamma$ induces. In other words, given a set of mechanistic rules $\gamma$, certain types of networks are more likely than others to be generated from the rules, but it is difficult to assign a probability to any given network, i.e., specify $P_{\mathscr{G}}(G = g \vert \gamma)$. The proposed framework identifies a collection of network properties (such as clustering and degree distribution), and their joint probability distribution, such that these elements fully characterize the networks generated by a given mechanistic model; we refer to this framework as Mechanistic to Probabilistic Model Conversion (MPMC). We consider a collection of network properties to be \emph{essential} for a model if the omission of any one property makes it no longer able to fully characterize the model. The goal of MPMC is to uncover the essential network properties and their joint probability distribution such that the probability to observe a network $g$ is identical whether the network is generated from the mechanistic model with rules $\gamma$ or is sampled from a probabilistic model with parameter $\omega$. 

There are advantages to being able to specify $P_{\mathscr{G}}(G = g \vert \gamma)$ as it enables investigators to perform statistical inference. In particular, the framework enables the investigation of whether a certain network property, such as clustering, is over- or under- represented in the generated networks compared to a reference model. There has been extensive research linking the presence or frequency of network properties to processes operating on the network, such as disease propagation. For example, a high clustering coefficient, i.e., an over representation of triangles, decreases the size of epidemics \citep{newman2003properties, MKKE05, miller2009percolation}. Degree assortativity has been associated with a network's resilience, a system's ability to function under abnormal conditions (such as node removal), and epidemic threshold, a metric characterizing whether an epidemic occurs \citep{MN02, newman2003mixing, boguna2003absence}. MPMC also enables investigation of additional questions by use of statistical hypothesis testing, such as whether two distinct rules, $\gamma^{1}$ and $\gamma^{2}$, generate identical networks, i.e., whether $P_{\mathscr{G}}(G = g \vert \gamma^{1}) = P_{\mathscr{G}}(G = g \vert \gamma^{2})$ for all $g \in \mathscr{G}$, or whether systems generated under two distinct rules have the same set of essential network properties. 

In the next section, we provide examples of mechanistic and probabilistic network models used to investigate the spread of HIV.  Section 3 provides current work on bridging these two models, limitations of this work, and a description of a recent class of probabilistic models, the Congruence Class Models (CCM), that overcomes some of these limitations and will later be used as part of MPMC. Section 4 provides details of the proposed MPMC framework, and Section 5 provides two examples of the MPMC framework using a mechanistic model designed to provide insight into the HIV epidemic. Section 6 discusses the proposed methods and suggests future research directions.

\section{HIV network models}\label{Examples}
HIV is a worldwide pandemic with an estimated 1.8 million new infections per year and is driven by biological and behavioral factors. Combining various strategies appears to be the most promising approach to HIV prevention \citep{kurth2011combination,  vermund2013can, bekker2012behavioral, buchbinder2010hiv, rotheram2009past, chang2013combination}, but it introduces financial, ethical, and logistical complexities that are best investigated using simulation-based approaches. 

Mechanistic and probabilistic network models have been used to investigate the spread of many communicable diseases \citep{keeling2005networks}. In particular, these models have been used to identify drivers of the HIV epidemic and to assess the impact of potential interventions, such as expanding access to antiretroviral therapy (ART) \citep{granich2009universal} and reducing concurrency \citep{MKHMW09}, i.e., individuals having multiple sexual partners simultaneously, on rates of new HIV infections. In the following two sections, we introduce examples of mechanistic and probabilistic network models that have had a significant role in guiding HIV interventions. The models introduced will be used to illustrate limitations of current approaches and to demonstrate the proposed MPMC framework.

\subsection{Mechanistic Models}

The family of mechanistic network models includes models that generate both idealized and realistic networks. Idealized network models investigate network structures or phenomena that occur across a range of settings, but typically are not sufficiently detailed for understanding specific systems, for planning interventions, or making policy recommendations for specific populations. For example, the small-world property refers to the idea that pairwise shortest path lengths are surprisingly small (logarithmic in $n$, the number of nodes) in most networks; this phenomenon allows infections to potentially reach any individual in the population over relatively short transmission chains. Another common phenomenon in social systems is the coalesence of influence to a few individuals \citep{price1976general, BA99}. This macroscopic phenomenon can emerge from a small collection of microscopic rules that encourage preferential attachment, the process where a new node introduced to the system adjoins to an existing node with a probability proportional to the number of edges the node already has, i.e., its degree. A characteristic feature of growing networks that are governed by preferential attachment is the emergence of a fat-tailed degree distribution. Preferential attachment rules were introduced in the model of Price for directed networks to study citation patterns of scientific papers \citep{price1976general}, and they were later introduced independently in a different formulation for undirected networks by Barab\'{a}si and Albert (BA) to describe a broad range of scientific and societal systems \citep{BA99}. The Price and BA models provide a mechanism to generate networks with a fat-tailed degree distribution, specifically a power-law degree distribution, where the probability, $P(k)$, that a vertex in the network has degree $k$, decays as a power-law $P(k) \sim k^{-\gamma}$. Subsequent research has shown that the BA model also produces non-random structure in other network properties besides degree distribution, including correlations between the degrees of connected nodes \citep{qu2015effects} and network clustering coefficient \citep{klemm2002growing}. The proposed MPMC framework can aid in such discoveries for the BA and other mechanistic models.

Realistic mechanistic network models attempt to include critical elements for a specific problem, and they enable researchers to guide trial designs and policy in the context of infectious diseases. In this paper, we apply our proposed framework to a mechanistic model developed by Kretzschmar and Morris--hereafter referred to as the KM model \citep{KM96con, MK97con}, which played a significant role in identifying intervention priorities by highlighting the potential impact of concurrency on epidemic spread in sub-Saharan African \citep{MGM07}. The model continues to be the building block of more recent realistic models to study HIV \citep{Pal12}. As it is believed that HIV epidemic in sub-Saharan Africa is driven by heterosexual relationships, the model only includes partnerships between people of the opposite sex, i.e., it is a model of a bipartite graph.

The network evolution under the KM model is based on individual-level stochastic rules for partnership formation and dissolution. The population is fixed and the relationships among the population form and dissolve over time. At each time $t$, an individual can form new partnerships, dissolve existing partnerships, or both. There are three key components governing the formation and dissolution of relationships: probability of pair formation ($p_f$), probability of pair separation ($p_s$), and a stochastic rule for partner mixing ($\phi$), which can depend on the properties of the nodes. (Section 3.2 provides further details on the three key components of the KM model.) The evolution of a network under the KM model is outlined below:

\begin{enumerate}
\item Let $g_t$ denote the network at time $t$. 
\item Repeat the following $T_1$ times:
\begin{enumerate}
\item Simulate a Bernoulli process where $X=1$ with probability $p_f$ and $X=0$ otherwise. 
\item If $X=1$: (i) Draw two unconnected individuals at random, one male, $i$, and one female, $j$; (ii) with probability $\phi(i,j)$ add edge $(i,j)$ to $g$, otherwise repeat (i) by redrawing two individuals at random.

\end{enumerate}
\item Every connected node pair splits up with probability $p_s$.
\end{enumerate}

\noindent The resulting network following these steps, represents the network at time $t+1$, denoted $g_{t+1}$. 

To use the KM model to simulate an HIV epidemic, one must specify an initial network at time $0$, denote this network as $g_0$. Once $g_0$ is specified, the steps outlined above can be used to generate networks at subsequent times. In the KM model, the network $g_0$ is generated by starting with an empty bipartite network with $n_1$ and $n_2$ nodes representing females and males, respectively, and then repeating the above steps a large number of times. This procedure is commonly referred to as a burn-in step. After completing this large number of iterations, the resulting network, $g_0$, is used at time $0$. The burn-in step ensures that the simulation of the HIV epidemic starts at the stationary state of the network generation process. In Section 5, we provide examples on how the MPMC framework can be used to derive a PMF for the stationary state of the process; therefore, one can sample a network $g_0$ from the stationary state instead of using the burn-in step described above. Note that in our paper, we focus only on the generation of the networks and not on modeling the HIV epidemic on the networks.

\subsection{Probabilistic model}

\begin{figure}[ht!]
\centering
\includegraphics[scale=0.5]{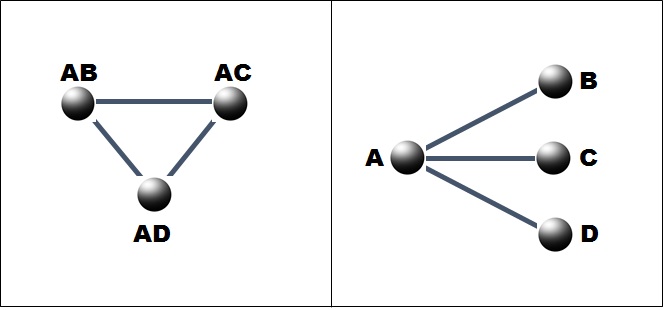}
\caption{Illustration of a Clique: An illustration of a clique $c$ is shown in the left panel. The corresponding subgraph $G_c$ is shown in the right panel.}
\label{clique}
\end{figure}

The breadth of probabilistic models is not as expansive as that of mechanistic models. The first probabilistic model was the Erd\H{o}s-R\'{e}nyi-Gilbert model \citep{ER60}. The use of the model on understanding epidemics is however limited as the network structure does not represent the structure of realistic populations. A common class of probabilistic network models that has been used to investigate HIV prevention interventions is the family of exponential random graph models (ERGM) \citep{FS86, robins2007introduction}. We provide technical details for deriving the ERGM probability distribution as these details will be important in understanding the limitations of ERGMs in generating networks identical to mechanistic models. In positing an ERGM, i.e., specifying $P_{\mathscr{G}}$, one proposes a dependency hypothesis that defines contingencies among the network edges, which are regarded as random variables; each potential edge, $E_{ij}$, has a corresponding random variable, denoted as $X_{ij}$. This hypothesis can be codified through the specification of a  dependence graph, denoted as $G_D = (V_D, E_D)$, on a population $V$.  The nodes of $G_D$ are tuples $(i,j)$, where $i,j \in V$.  An edge in $G_D$ is represented as a pair of tuples, i.e., $\{(i,j), (k,m)\}$ where $i,j,k,m \in V$. Here $\{(i,j), (k,m)\}$ is an edge in $G_D$ if and only if edges $(i,j)$ and $(k,m)$ are conditionally dependent given information on all other potential edges, that is, the probability of the edge $(i,j)$ existing in a graph depends on the presence of edge $(k,m)$. Let $C$ denote the set of cliques in $G_D$. Let $G_c$ be the graph formed by the collection of all edges denoted by the nodes of $c \in C$; Figure~\ref{clique} provides an illustration of a clique $c$ and corresponding subgraph $G_c$.

The Hammersley-Clifford theorem states that $P_{\mathscr{G}}$ is a Gibbs distribution that can be factored over $G_D$, conditional on $P_{\mathscr{G}}$ being a positive distribution, i.e, $P_{\mathscr{G}}(G = g) > 0$ for all $g \in \mathscr{G}$  \citep{besag1974spatial}. Therefore, 

\begin{equation} \label{eq:Hammersley_Clifford}
P_{\mathscr{G}}(G = g) = \frac{1}{Z} \prod_{c \in C} \psi_c (X_c), 
\end{equation}

\noindent where $\psi_c$ is a function over sets of variables $X_c$ associated with clique c in $G_D$, and $Z$ is a normalizing constant. 
As Equation~\ref{eq:Hammersley_Clifford} does not provide a unique distribution, additional constraints are necessary. A natural set of constraints is to assign the probability of observing $G_c$ for each $c \in C$. These constraints control the probability that a subgraph, where the presence of all the edges are dependent on each other, is observed; they are represented in Equation~\ref{eq:mean_constraints}:

\begin{equation} \label{eq:mean_constraints}
\sum_{g \in \mathscr{G}} I_{G_c \subseteq g}P_{\mathscr{G}}(G = g) = P_{\mathscr{G}}(I_{G_c \subseteq g}),
\end{equation}

\noindent where $I_{G_c}$ is the indicator function that $G_c$ is a subgraph of $g$ and $P_{\mathscr{G}}(I_{G_c \subseteq g})$ is the probability that needs to be specified. 

As all subgraphs of $G_c$ are associated with a clique in $C$, they too would be the subject of a constraint. Even with these constraints, $P_{\mathscr{G}}$ is not uniquely defined. A potential probability distribution is one that maximizes the Shannon entropy subject only to constraints represented in Equation~\ref{eq:mean_constraints}; the maximum entropy principle is conceptually powerful and finds numerous applications in science, and in particular physics \citep{presse2013principles}. The maximum entropy distribution best represents the current state of knowledge of a system, while assuming maximal ignorance  about the distribution other than those imposed by Equation~\ref{eq:mean_constraints} \citep{MN10, cimini2019statistical}. This approach leads to the following distribution: 

\begin{equation} \label{eq:max_entropy_distr}
P_{\mathscr{G}}(G = g) = \frac{1}{Z} \prod_{c \in C} \exp(\omega_{G_c} I_{G_c \subseteq g}),
\end{equation}

\noindent where $\omega_{G_c}$ is a parameter used to fix the mean probability of observing $G_c$, i.e., specify $P_{\mathscr{G}}(I_{G_c \subseteq g})$. Therefore, 

\begin{equation} \label{eq:potential_func}
\psi_c (X_c) = \exp(\omega_{G_c} I_{G_c \subseteq g}). 
\end{equation}

As the distribution specified in Equation~\ref{eq:max_entropy_distr} has a large number of parameters, $\{\omega\}$, one simplifies the model by imposing a homogeneity assumption by equating parameters when they refer to the same type of configuration, e.g., triangles. The resulting PMF presented below is the standard form for ERGMs:

\begin{equation} \label{eq:ERGM}
P_{\mathscr{G}}(G = g | \omega) \propto \exp(\omega^T \eta(g)),
\end{equation}

\noindent where $\omega$ is a (column) vector of model parameters associated with the specified network properties and $\eta(g)$ denotes the vector of counts for the network configuration associated with the cliques in $G_D$ (also referred to as sufficient network statistics for the ERGM), i.e., $\eta: \mathscr{G_n} \rightarrow \mathbb{R}^p$, where $p$ is the length of the vector. As referred by Cimini et al. \citep{cimini2019statistical}, ERGMs is an example of a canonical approach, that is an approach where networks are  generated to have network features that match the observed network in expectation; this is in contrast with microcanonical approaches which generate networks that exactly match observed network properties, for example the configuration model \citep{molloy1995critical, MN10}.  

\section{Previous Work: Synergizing Mechanistic and Probabilistic Models}

The connection between a PMF and a set of mechanistic rules has been studied for particular models, though presented in a different context.  In particular, the probability distribution of networks generated by particular mechanistic rules can be represented as an ERGM. In a few settings, a dependence graph can be created based on the assumptions that underpin the mechanistic rules. For example, if the mechanistic rule is that individuals form edges with a fixed probability $p$ and independent of all other edges, i.e., the Bernoulli assumption holds, the dependence graph is the empty graph. Therefore, each clique in $G_D$ consists of one node. This leads to an ERGM with one sufficient statistic, the number of edges, once imposing the homogeneity assumption that all parameters associated with the same network configuration--in our case, a single edge--are equal; we obtain the following ERGM Equation~\ref{eq:ERGM_edge}.

\begin{equation} \label{eq:ERGM_edge}
P_{\mathscr{G}}(G = g | \omega) \propto \exp(\omega \eta(g)),
\end{equation}

\noindent where $\eta(g)$ represents the number of edges in $g$ and $\omega = \log(p/1-p)$.

In the following, we highlight some limitations of representing mechanistic models using ERGMs and present a recent network model that overcomes some of these limitations.

\subsection{Limitations of ERGMs}

In developing the PMF for ERGMs there are two critical requirements. The first is that the dependence graph, $G_D$, is not complete. A complete dependence graph results in $2^{\binom{n}{2}}$ cliques, which leads to a large number of parameters in Equations~\ref{eq:max_entropy_distr} and \ref{eq:ERGM} as well as identifiability issues; dense dependence graphs may also be problematic for a similar reason. The second requirement is that Equation~\ref{eq:mean_constraints} represents the only constraints about the system, that is, there are only constraints regarding the mean of network configuration counts; this excludes including information on the second or third moments on network configurations. For instance Equation~\ref{eq:mean_constraints} does not allow specifying uncertainty in those counts (measurement error) and variability around those counts (due to the stochastic nature of the mechanistic rules).

As there is a tendency for network properties to exhibit sharp threshold effects, slight errors in estimation of network configuration counts can have a major impact on beliefs about the overall structure of the network. Therefore, it is essential for researchers to utilize knowledge about the mechanistic model, and in particular, the variability of network property estimates. Though ERGMs are quite flexible, these two requirements result in challenges to modeling mechanistic network models using ERGMs. The challenges are demonstrated through an investigation of the BA and KM mechanistic network models. These simple demonstrations illustrate the limitations imposed by these two requirements and the need for a more flexible probabilistic network model. 

\subsubsection{BA model}

The BA model can be initiated with a small seed graph, which  grows  by the addition of new nodes one at a time. (The model can be  modified in various ways, but we consider only the original version of the model.) Nodes and edges, once introduced, are never deleted. Each new node forms exactly $m_0$ new edges with existing nodes based on preferential attachment rules. Specifically, the probability that a new node $i$ connects to node $j$ is the following:

\begin{equation} \label{eq:random_mixing_f_prob}
p_j = \frac{d_j}{\sum_k d_k}.
\end{equation} 

\noindent Therefore, when determining the probability of an edge between the new node $i$ and an existing node $j$, one needs to know the degree of $j$ as well as the degrees of all other nodes in the network. These requirements necessitate that the dependence graph is complete--violating the first ERGM requirement. 

\begin{figure}[ht!]
\centering
\includegraphics[scale=0.75]{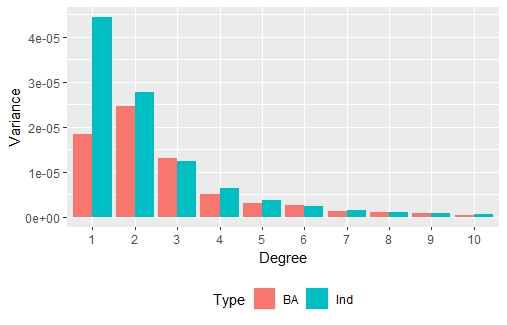}
\caption{Comparison between BA model and Independence: A comparison between the variance of the degree distribution generated from the BA model (n = 5000, m = 1) and the variance of the multinomial distribution where the parameters are specified as the mean associated with the BA model.}
\label{BA_example}
\end{figure}

As mentioned above, the degree distribution of the BA model follows the power law. In addition, the BA model constrains the variability in the degree distribution compared to the maximum entropy probability distribution where the only constraints are the means of the degree distribution associated with the BA model; this maximum entropy probability distribution is the multinomial distribution \citep{kapur1989maximum}. Figure~\ref{BA_example} provides a comparison between the variance of the degree distribution generated from a BA model where $m_0 = 1$ and the variance of the multinomial distribution where the parameters are specified as the means of the degree distribution associated with the BA model. The degree distribution variance for the BA model is smaller than the variance of the multinomial distribution. Therefore, the BA model imposes constrains beyond just the mean proportions, such as the last node added must have degree equal to $m_0$; this means that the BA model also violates the second ERGM requirement.

\subsubsection{KM model}

In the KM model, the probability of an edge forming between nodes $i$ and $j$ depends on the mixing function $\phi$ and the number of edges in the network, which is controlled by $T_1$. Similar to the BA model, the KM model violates both of the ERGM requirements. First, the formation of any edge is dependent on the existence of all the other edges due to the inclusion of the random process associated with parameter $T_1$. Therefore, the dependence graph is a complete graph. 

Similar to the BA model, the KM model constrains the variability of network properties, in particular number of edges, compared to the maximum entropy probability distribution, where the only constraints are the means of essential network properties. To illustrate this point and the limitation of ERGMs to capture the KM model, we: (1) simulate $k$ networks, $\{g_{M_1},\cdots,g_{M_k}\}$, using a specification of the KM model; (2) sample $k$ networks, $\{g_{S_1},\cdots,g_{S_k}\}$, from an ERGM that includes the essential network properties of the chosen KM model; and (3) compare $\{g_{M_1},\cdots,g_{M_k}\}$ to $\{g_{S_1},\cdots,g_{S_k}\}$. The following provides additional details on each step.\\

\noindent \underline{\textit{Step 1: Simulate from KM}}\\

We investigate a simple specification of the KM model, pure random mixing, and we use identical parameter values as the authors of the KM model when it was first proposed \citep{KM96con}. In the pure random mixing setting, there exists no preference for nodes to form edges based on the covariates of the nodes. The $\phi$ function for this setting is the following:

\begin{equation} \label{eq:random_mixing_f_prob_2}
\phi(i,j) = \left\{ \begin{gathered}
1 \mbox{ if } k_i < d_{m} \mbox{ and } k_j < d_{m} \hfill \\
0 \mbox{ else, } \hfill \\
\end{gathered} \right.
\end{equation}

\noindent where $k_i$ and $k_j$ are the current degrees of nodes $i$ and $j$. The following parameters values were used in the original model: $n_1 = n_2 = 1000$, $p_{f} = 0.01$, $p_{s} = 0.005$, $T_1 = (n_1 + n_2)/2 - \vert g \vert$, and $d_{m} = 10$.\\

\noindent \underline{\textit{Step 2: Simulate from ERGM}}\\

In Section 5, we provide evidence that the essential network properties for the KM model for pure random mixing only includes number of edges. Therefore, we look at ERGMs where the number of edges is included as the only network statistic as we know other properties are not relevant.  We investigate two ERGMs: (1) one that includes number of edges, and (2) one that includes number of edges and a constraint that the number of edges cannot exceed $(n_1 + n_2)/2$, a constraint that is implicit in the KM model.\\

\noindent \underline{\textit{Step 3: Comparison}}\\

\begin{figure}[ht!]
\centering
\includegraphics[scale=0.4]{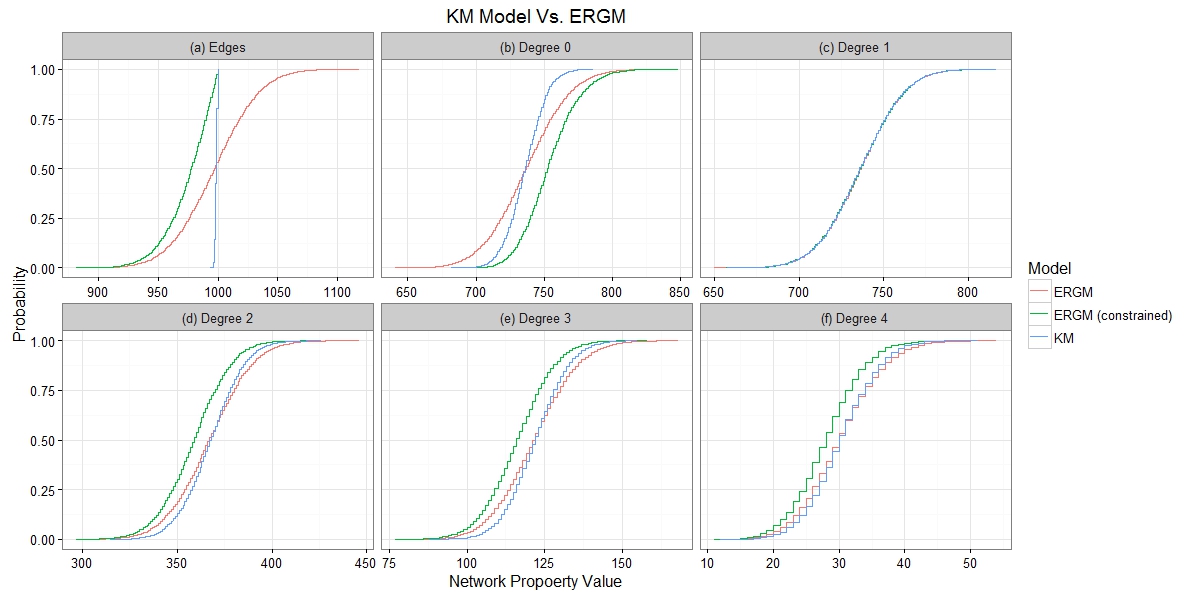}
\caption{Comparison between KM model and ERGMs: A comparison of the number of edges and number of nodes of specified degree across the network collection for the KM model and ERGMs. Panel (a) depicts the CDF for the number of edges. Panels (b)-(f) depict the CDF for the number of nodes with degrees $\{0,1,\cdots, 4\}$. The blue lines depict the CDFs for the KM model, and the red and green lines depict the CDFs for the ERGMs with and without the constraint on the number of edges, respectively. Because the CDFs generally do not match, the specified ERGMs are not able to fully capture the network structure generated by the KM model.}
\label{RM_model_ERGM_comp}
\end{figure}

We compare the cumulative density functions (CDFs) of the two collections of networks, $\{g_{M_1},\cdots,g_{M_k}\}$ and $\{g_{S_1},\cdots,g_{S_k}\}$, on network properties that consist of number of edges and number of individuals with degrees $\{0,1,\cdots, 4\}$ (CDFs for degrees 5-10 are not shown here as few nodes had degrees in this range). The blue lines in Figure \ref{RM_model_ERGM_comp} depict the CDFs of the network properties for the $k$ networks generated by the KM mechanistic model. The red and green lines depict the CDFs of the network properties for the $k$ networks sampled from the ERGM with and without the constraint on the number of edges, respectively. The CDF associated with the KM model in blue is significantly steeper than the ERGMs for number of edges and number of nodes of degree 0, and only slightly steeper for the remaining degrees. The steeper CDFs for the KM model compared to the ERGM models indicates that the mechanistic model imposes constraints on the variability of the examined network properties compared to ERGMs. 

\subsection{Congruent Class Model}\label{Methods}

The limitations of ERGMs to adequately model the BA and KM mechanistic models illustrates the need for greater flexibility in the modeling of network properties. To overcome some of these limitations, the proposed MPMC framework uses the Congruent Class Model (CCM) \citep{goyal2014sampling}, which allows for greater flexibility in specifying the functional form of the probability distributions associated with network properties. 

The CCM partitions the space of graphs on $n$ nodes, $\mathscr{G}$, such that all graphs within a partition have the same values for the network properties of interest; these partitions are referred to as congruence classes. For example, one congruence class might correspond to all graphs with 50 closed triads, another congruence class to graphs with 51 closed triads, and so on. Therefore, a congruence class is defined as $c_{x} = \{g : \eta(g) = x, g \in \mathscr{G_n}\}$, where $\eta(g)$ denotes the value of the properties used to define the congruence classes for $g \in \mathscr{G}$. The number of networks in $c_{x}$ is denoted as $\vert c_{x} \vert$, which is referred to as the volume factor \citep{Shalizi13}. The probability distribution on $\mathscr{G_n}$ for the CCM is based on specifying $P_{\mathscr{C}}$, the probability mass function for the congruence classes defined by the essential network properties; $P_{\mathscr{C}}(x)$ is the total probability of all networks that are elements in $c_{x}$:

\begin{equation} \label{eq:networkprob_static_pc}
P_{\mathscr{C}}(x) = \sum_{g \in c_{x}} P_{\mathscr{G}}(g).
\end{equation}

Since the congruence classes represent the partition of the space $\mathscr{G}$ based on essential network properties, two networks within a congruence class must have the same probabilities of being observed. Therefore, the probability distribution on $\mathscr{G}$ for the CCM is the following:

\begin{equation} \label{eq:networkprob_static}
P_{\mathscr{G}}(G = g) \: \propto \: \left(\frac{1}{\vert c_{\eta(g)} \vert} \right) P_{\mathscr{C}}(c_{\eta(g)}).
\end{equation}

The flexibility of the CCM results from allowing a broad range of models, including both parametric and non-parametric models, to be used to assign the PMF on the defined congruence classes, $P_{\mathscr{C}}$. CCMs generalize many common network models including the Erd\H{o}s-R\'{e}nyi-Gilbert model, stochastic block model, and ERGMs when nodal attributes are discrete. For instance, to specify a probability distribution identical to the Erd\H{o}s-R\'{e}nyi-Gilbert model, one would set $P_\mathscr{C}(c_{\eta(g)})$ equal to the following: 

\begin{equation} \label{eq:ER_CCM_prob}
P_\mathscr{C}(c_{\eta(g)}) = p^{\eta(g)}(1-p)^{{n \choose 2} - \eta(g)} {{n(n-1)/2} \choose \eta(g)},
\end{equation}

\noindent where $\eta(g)$ represents the number of edges in $g$. 

The CCM and ERGM are similar in that both models characterize networks by their network properties. The difference lies in the specification of the probability distribution of the network properties. For ERGMs, the investigator cannot specify a probability distribution for essential network properties; the distribution for each essential network property is completely specified by a single value, referred to as a sufficient statistic. This is analogous to the binomial distribution in elementary probability and statistics, where one only needs to estimate the probability of a success to specify the entire probability distribution. By contrast, the CCM model does not place restrictions on the joint probability distribution of network properties; the probability mass function on congruence classes, $P_\mathscr{C}$, is flexible. This flexibility allows the model to handle the complexity that can arise in mechanistic models. 

A complication for CCM is that a calculation for each set of network properties needs to be derived in order to generate samples from the CCM; calculations for network properties that include network density, degree distribution, mixing by covariate and degree, and triangles (closed triads) have been evaluated for uni-modal graphs \citep{goyal2014sampling}; these calculations (when appropriate) have been expanded to bipartite graphs \citep{goyal2018inference}. 

\section{Framework}\label{Framework}

\begin{figure}[ht!]
\centering
\includegraphics[scale=0.5]{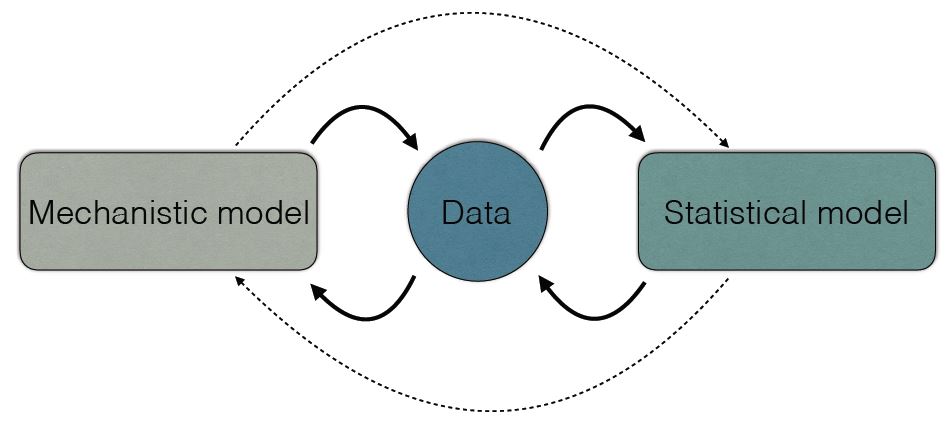}
\caption{Conceptual Illustration of the Conversion Framework: The Mechanistic to Probabilistic Model Conversion (MPMC) framework first generates network realizations from a mechanistic network model, then uses an iterative algorithm to capture the joint distribution of so-called essential network properties, and then uses those statistics in a statistical model (in our case, the congruent class model) to generate networks that are indistinguishable from those generated with the mechanistic model.}
\label{MtSMC}
\end{figure}

To characterize a mechanistic model with microscopic rules $\gamma$, the MPMC framework needs to identify the essential network properties along with their joint probability distribution. The general MPMC framework is an iterative algorithm where a collection of network properties is proposed as the essential network properties, their suitability is assessed and, depending on the conclusion, either a new collection of network properties is proposed or the algorithm terminates. Figure~\ref{MtSMC} provides a conceptual illustration of the MPMC framework and an outline of the conversion framework is as follows:

\begin{enumerate}
\item \textit{Simulate the mechanistic model:} Generate a collection of networks, $\{g_{M_1},\cdots,g_{M_k}\}$, based on simulating the mechanistic model $k$ times.
\item \textit{Propose essential network property candidates: } Based on subject matter knowledge, conceptual knowledge of the mechanisms, and previous iterations of the algorithm, propose a collection of network properties, defined by the function $\eta$, as the essential network properties of the mechanistic model.
\item \textit{Estimate the joint probably distribution of essential network properties: } Estimate the joint probability distribution, $P_{\mathscr{C}}$, of the candidate essential network properties, defined by $\eta$, based on the observed simulated networks $\{g_{M_1},\cdots,g_{M_k}\}$. In high dimensions, i.e., settings where a large number of network properties is being considered, density estimation is a non-trivial problem. However, given the generic nature of the problem, there exists a vast literature on methods for density estimation in this setting \citep{silverman1986density, scott2015multivariate}.
\item \textit{Sample networks: } Sample networks, $\{g_{S_1},\cdots,g_{S_k}\}$, based on a CCM with the estimated joint probability distribution $P(s_1, ..., s_j)$. 
\item \textit{Compare networks: } Statistically compare the probability distribution of the two collections of networks, $\{g_{M_1},\cdots,g_{M_k}\}$ and $\{g_{S_1},\cdots,g_{S_k}\}$, on a large set of network properties defined by $\eta'$ not contained in the set defined by $\eta$.
\item \textit{Iterate: } If statistical tests do not reject the hypothesis that the probability distributions on each of the network properties defined by $\eta'$ differ between $\{g_{M_1},\cdots,g_{M_k}\}$ and $\{g_{S_1},\cdots,g_{S_k}\}$, then accept the properties defined by $\eta$ as the essential network properties, such that their joint probability distribution $P_{\mathscr{C}}$ fully characterizes the network properties induced by the mechanistic model. Otherwise repeat steps 2-6.
\end{enumerate}

\section{Application}\label{Application}

In this section we investigate the KM model described in Section 2. Specifically, we investigate two rules for partner mixing: serial monogamy and pure random mixing. In neither one of these settings is it straightforward to understand the implications of the mechanistic rules of the KM model on the properties of the induced networks. We use identical parameter values as the authors of the KM model when it was first proposed \citep{KM96con} and shown in Section 2.

\subsection{Pure Random Mixing}

To characterize the pure random mixing setting of the KM model, i.e., to identify the essential network properties of the mechanistic model along with their joint probability distribution, we follow the steps of the MPMC framework outlined in Section 4. As described in Section 2, in the random mixing setting, there exists no preference for individuals to form relationships based on the degree.

\begin{enumerate}
\item \textit{Simulate the mechanistic model:} Let $\gamma_r$ denote the microscopic rules associated with the pure random mixing setting for the KM model. We simulate $k$ networks, $\{g_{M_1},\cdots,g_{M_k}\}$, based on $\gamma_r$.

\item \textit{Propose essential network property candidates: } Based on Figure \ref{RM_model_ERGM_comp} it may appear that modeling the degree distribution would be necessary, however, we propose modeling only number of edges as the essential network property of the KM model. Let $X_E^{\gamma_r}$ represent the random variable for the number of edges in a network generated with $\gamma_r$.

\item \textit{Estimate the joint probability distribution of essential network properties: } Let
$P_E^{\gamma_r}$ denote the probability mass function for $X_E^{\gamma_r}$. From the blue line in panel (a) of Figure \ref{RM_model_ERGM_comp}, it appears that the distribution $P_E^{\gamma_r}$ does not follow any common distribution; therefore, we estimate $P_E^{\gamma_r}$, denoted as $\hat{P}_{E}^{\gamma_r}$, by letting $\hat{P}_{E}^{\gamma_r}( X_E^{\gamma_r} = x)$ equal the fraction of the $k$ generated networks that have $x$ edges, i.e., $\hat{P}_{E}^{\gamma_r}( X_E^{\gamma_r} = x) = \frac{1}{k} \sum_{i=1}^k I_{\eta(g_{M_i}) = x}$.

\item \textit{Sample networks: } We sample networks based on the following probability mass function: 

\begin{equation} \label{eq:networkprob_static_RM}
P_{\mathscr{G}}(G = g \vert P_E^{\gamma_r}) \: \propto \: \left(\frac{1}{\vert c_{\eta(g)} \vert} \right) P_{\mathscr{C}_r}(c_{\eta(g)}),
\end{equation}

\noindent where $P_{\mathscr{C}_r}(c_{\eta(g)}) = \hat{P}_{E}^{\gamma_r}( X_E^{\gamma_r} =  \eta(g))$ and $\eta(g)$ is the number of edges in $g$. 

\item \textit{Compare networks: } 
Figures \ref{RM_res_deg_dist} and \ref{RM_res_net_prop} compare the networks generated from the KM model and those sampled from the CCM based on equation \ref{eq:networkprob_static_RM} on a large set of network properties which consists of the number of edges, number of nodes of degree 0-4 (nodes of higher degree were extremely rare), betweenness centrality (max and mean across all nodes), degree correlation, eigenvalue centrality (max and mean across all nodes), and number of K-stars (1-3); detailed descriptions of the metrics are available in \citep{wasserman1994social} and \citep{MN10}. Based on the Kolmogorov--Smirnov test, one cannot reject the hypothesis that the network property distributions are identical (the $p$-values ranged from $0.23$ to 1 across all of the network properties) \citep{arnold2011nonparametric}. 

\begin{figure}[ht!]
\centering
\includegraphics[scale=0.5]{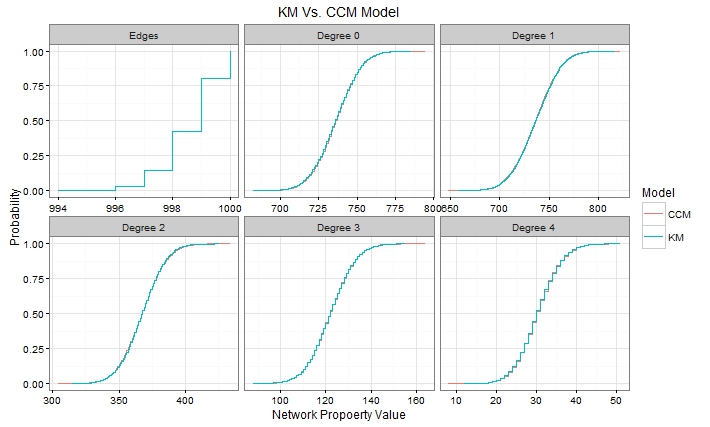}
\caption{Comparison Between KM Model and CCM on the Number of Edges and Degree Distribution: A comparison of the number of edges and number of nodes of specified degree across the network collection for the KM model and CCM ERGMs. Panel (a) depicts the CDF for the number of edges. Panels (b)-(f) depict the CDF for the number of nodes with degrees $\{0,1,\cdots, 4\}$. The red lines depict the CDFs for the KM model, and the blue lines depict the CDFs for the CCM. Because the CDFs match perfectly, the specified CCM appears to be able to capture the network structure generated by the KM model.}
\label{RM_res_deg_dist}
\end{figure}

\begin{figure}[ht!]
\centering
\includegraphics[scale=0.5]{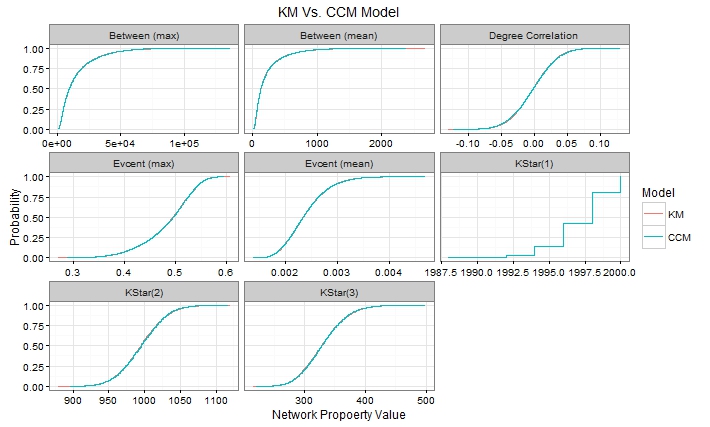}
\caption{Comparison Between KM Model and CCM on Higher Order Properties: A comparison of centrality measures (betweenness and eigenvector), degree correlation, and number of $k$-stars across the network collection for KM model and CCM. Panels (a) and (b) depict the CDF for the max and mean betweenness centrality. Panel (c) depict the CDF for the degree correlation. Panels (d) and (e) depict the CDF for the max and mean eigenvector centrality. Panels (f)-(h) depict the CDF for the number of k-stars with k equal to $1$, $2$, and $3$. The red lines depict the CDFs for the KM model, and the blue lines depict the CDFs for the CCM. Because the CDFs match perfectly, the specified CCM appears to be able to capture the network structure generated by the KM model.}
\label{RM_res_net_prop}
\end{figure}

\item \textit{Iterate: } Based on the Kolmogorov--Smirnov tests, we conclude that the number of edges is the only essential network property, and the probability distribution  in equation \ref{eq:networkprob_static_RM} fully characterizes the mechanistic random mixing KM model. 

\end{enumerate} 

\subsection{Serial monogamy}

In the serial monogamy setting, individuals are restricted from having more than one partner at the same time. In the paper by \citep{KM96con}, the $\phi$ function for this setting is the following:

\begin{equation} \label{eq:serial_mono_f_prob}
\phi(x,y) = \left\{ \begin{gathered}
1 \mbox{ if } k_i = k_j = 0 \hfill \\
0 \mbox{ else. } \hfill \\
\end{gathered} \right.
\end{equation}

\noindent For the remaining parameters, we use identical values as the authors of the KM model when it was first proposed \citep{KM96con} (see Section 2).

As in the previous example, to characterize the serial monogamy setting of the KM model, i.e., identify the essential network properties of the mechanistic model along with their joint probability distribution, we follow the steps of the MPMC framework outlined in Section 4.

\begin{enumerate}
\item \textit{Simulate the mechanistic model:} Let $\gamma_s$ denote the microscopic rules associated with the serial monogamy setting for the KM model. We simulate $k$ networks, $\{g_{M_1},\cdots,g_{M_k}\}$, based on $\gamma _s$.

\item \textit{Propose essential network property candidates: } Our candidate collection of essential network properties include only the number of individuals with degree 0. Let $X_{D_0}^{\gamma_r}$ represent the random variable for the number degree 0 nodes generated with $\gamma _s$.

\item \textit{Estimate the joint probability distribution of essential network properties: } Let
$P_{D_0}^{\gamma_s}$ denote the probability mass function for $X_{D_0}^{\gamma_s}$. We estimate $P_{D_0}^{\gamma_s}$, denoted as $\hat{P}_{D_0}^{\gamma_s}$, by letting $\hat{P}_{D_0}^{\gamma_s}( X_{D_0}^{\gamma_s} = x)$ equal the fraction of the $k$ generated networks that have $x$ individuals with degree 0.

\item \textit{Sample networks: } We sample networks based on the following probability mass function: 

\begin{equation} \label{eq:networkprob_static_MG}
P_{\mathscr{G}}(G = g \vert \hat{P}_{D_0}^{\gamma_s}( X_{D_0}^{\gamma_s} = x)) \: \propto \: \left(\frac{1}{\vert c_{\eta(g)} \vert} \right) P_{\mathscr{C}}^{\gamma_s}(c_{\eta(g)}),
\end{equation}

\noindent where 
$P_{\mathscr{C}}^{\gamma_s}(c_{\eta(g)}) = \hat{P}_{D_0}^{\gamma_s}( X_{D_0}^{\gamma_s} = \eta(g))$ and $\eta(g)$ is the number of nodes with degree 0.

\item \textit{Compare networks: } We compare the networks generated from the KM model and those generated from the CCM based on equation \ref{eq:networkprob_static_MG} on a large set of network properties which consists of the number of edges, number of nodes of degree 0 and 1 (nodes of higher degree are not compatible with the monogamy model), and eigenvalue centrality (max, mean, median, and min across all nodes). Based on the Kolmogorov--Smirnov test, one cannot reject the hypothesis that the network property distributions are identical (the $p$-values ranged from 0.96 to 1 across all of the network properties).

\item \textit{Iterate: } Based on the Kolmogorov--Smirnov tests, we conclude that the number of individuals with degree 0 is the only essential network property, and the probability distribution $P_{D_0}^{\gamma_s}$ fully characterizes the serial monogamy KM mechanistic model. 

\end{enumerate} 

Note that as individuals either have degree 0 or 1, it would be equivalent to use the number of individuals of degree 1 as our essential network property.

\section{Discussion}

In this paper, we have proposed the Mechanistic to Probabilistic Model Conversion (MPMC) framework for first learning the joint distribution of essential network properties of a mechanistic network model and then using a probabilistic model, the congruence class model (CCM), to generate networks that are indistinguishable from those generated by the original mechanistic model. An illustration of two examples of mechanistic models, which are based on relatively simple rules, demonstrate the complexity that can result from mechanistic models. This complexity exposes limitations on representing mechanistic models using probabilistic models, in particular ERGMs. Therefore, the CCM was used as the probabilistic model as it overcomes some of these limitations. 

In Section 2, we highlight two requirements of ERGMs: (1) the dependence graph, $G_D$, is not complete and (2) Equation~\ref{eq:mean_constraints} represents the only constraints about the system. The two examples violate both requirements. However, in general, it is not straightforward to assess if a mechanistic model violates these requirements. For example, there is a modification to the BA where a new node chooses an existing node and then selects a neighbor of that existing node at random \citep{saramaki2004scale}. Only after careful consideration it is clear that this modified BA model violates the requirement that the dependence graph not be complete.

We identified two areas of promising research. The first is addressing the other direction of the MPMC framework, i.e., proposing mechanistic models that are consistent with a probability distribution; see \citep{Chen2018arXiv} for initial work in this area. The second is understanding the flexibility that a probabilistic model must have to represent all mechanistic models or particular classes of mechanistic models.

Finally, we have kept the examples simple to demonstrate a proof of concept of the framework, and we acknowledge that much additional work is needed to more fully bridge the two approaches. In any case, the proposed framework provides a novel method that has potential for investigators to gain new insights by synergizing the two approaches to network modeling. 

\section*{Acknowledgements}

This research is supported by the following grants from the National Institutes of Health: R37AI51164, R01AI138901. We would like to thank Victor De Gruttola for his useful ideas and comments.

\bibliographystyle{unsrt}  
\bibliography{paper2_bib}

\end{document}